\documentclass{ctr_summer}

\usepackage{ctrfont}
\usepackage{natbib}
\usepackage{undertilde}
\usepackage{color}

\usepackage{graphicx}

\usepackage{amsmath,rotating}
\usepackage{dashrule}
\usepackage{undertilde}

\usepackage{url}

\usepackage{color}

\usepackage{bbold}
\renewcommand{\d}[1]{\frac{\partial #1}{\partial t}}
\renewcommand{\b}[1]{\boldsymbol{#1}}
\newcommand{\T}[0]{^H}

\title{Data-driven linear analysis of turbulent flows\\via nonlinearity-subtracted dynamic\\mode decomposition}

\shorttitle{Data-driven linear analysis of turbulent flows}

\author{B.~Herrmann\footnote[1]{Department of Mechanical and Metallurgical Engineering, Department of Hydraulic and Environmental Engineering, Pontificia Universidad Católica de Chile, Chile}, K.~Cao, C.~A.~Gonzalez, S.~L.~Brunton\footnote[2]{Department of Mechanical Engineering, University of Washington} \and B.~J.~McKeon}

\shortauthor{Herrmann et~al.}

\begin{document}

\setcounter{page}{1}
\pagenumbering{gobble}
\maketitle

Mean-flow-based linear analyses of turbulent flows, such as resolvent analysis, provide valuable insight about flow structures and their dynamics that has been widely leveraged to model, control and understand the underlying flow physics. However, these analyses are computationally expensive for flows over complex geometries and require the use of specialized codes that are typically only available in research environments. On the other hand, data-driven modal decompositions, such as the dynamic mode decomposition (DMD), identify turbulent flow structures that, although statistically relevant, do not provide insight into the physical mechanisms driving their dynamics. Here we introduce a novel data-driven method—nonlinearity-subtracted DMD (NSDMD)—that leverages knowledge of the structure of the Navier–Stokes equations to ensure that the learned operator is a low-rank approximation of the underlying mean-flow-linearized dynamics. Specifically, the method uses snapshots of the nonlinear terms in the perturbation equations to explicitly account for the contribution of the nonlinear forcing to the dynamics. We demonstrate the use of NSDMD to perform data-driven resolvent analysis on direct numerical simulation (DNS) and large-eddy simulation (LES) datasets, starting with a minimal channel flow and scaling up to the flow over a full aircraft model. As a result, NSDMD allows performing linear analyses of turbulent flows as a post-processing step on simulation data obtained with any available high-fidelity computational fluid dynamics (CFD) code.\\

\hrule

\section{Introduction}
  
Linear analysis tools provide valuable insight about dominant flow structures and their dynamics that has been widely leveraged to model and understand the underlying flow physics~\citep{jovanovic2021arfm}. For the case of turbulent flows, mean-flow-based analyses have been leveraged to improve our understanding of the linear amplification mechanisms sustaining turbulence~\citep{mckeon2017jfm}.
 During the last decade, resolvent analysis has been extensively used for understanding, predicting and controlling laminar and turbulent flows~\citep{mckeon2017jfm,herrmann2023jfm,rolandi2024tcfd}.

Unfortunately, despite its relevance for applications, mean-flow-based linear analysis of flows in complex geometries with three inhomogeneous directions is still not commonplace in industrial settings. The main reasons for this are the high computational cost and large memory allocation required to handle a very large operator.
 Although enormous progress has been made in addressing these challenges using matrix-free time stepping~\citep{frantz2023amr} and randomized linear algebra~\citep{ribeiro2020prf}, global resolvent analysis still requires the use of specialized in-house codes that are only available in research environments.

In contrast, the unprecedented availability of high-fidelity data from numerical simulations and experimental measurements of fluid flows has led to the development and application of many data-driven modal decompositions~\citep{taira2017aiaa}.
The dynamic mode decomposition (DMD) is a particularly relevant technique that has been consolidated as a basic tool for data-driven analysis of fluid flows, allowing us to simultaneously identify relevant flow structures and fit a linear model to their dynamics from time-resolved measurements~\citep{schmid2010jfm,schmid2022arfm}.
Physics-informed DMD (piDMD) is a recently developed extension that allows the integration of known physical principles, such as symmetries, invariances and conservation laws, into DMD models~\citep{baddoo2023prsa}. 
However, for a turbulent flow, so far all of these data-driven modal decompositions identify flow structures that hold no direct relation to the governing equations and, therefore, do not provide physical insight into the mechanisms sustaining turbulence.

However, we recently formulated a data-driven approach to perform resolvent analysis of laminar flows~\citep{herrmann2021jfm}. Data-driven resolvent analysis (DDRA) leverages DMD to compute resolvent modes and gains directly from time-resolved measurement data, without requiring the governing equations~\citep{herrmann2021jfm}.
Note that DDRA was developed for linear stable systems; therefore, its application to turbulent flows requires separating the nonlinear contributions to the dynamics in the data. This separation can be attempted with methods that simultaneously fit the linear and nonlinear contributions to the dynamics, such as the linear and nonlinear disambiguation optimization (LANDO)~\citep{baddoo2022prsa} and operator inference~\citep{kramer2024arfm}. However, although these methods identify nonlinear models that can produce accurate short-term predictions, their linearization does not necessarily agree with the linearization of the original system.

\begin{figure}
    \centering
    \includegraphics[width=1\linewidth]{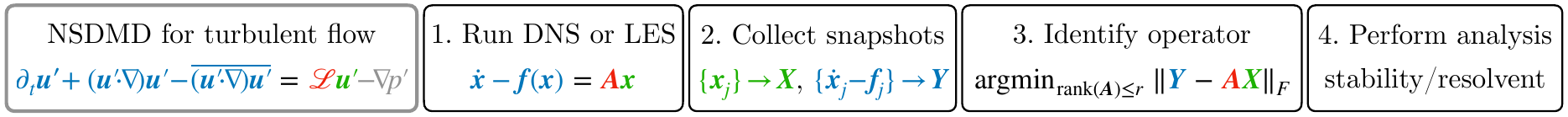}
    \caption{Schematic of the procedure followed by NSDMD for turbulent flows.}
    \label{fig_nsdmd}
\end{figure}

In this work, we propose a novel data-driven method—nonlinearity-subtracted DMD (NSDMD)—that focuses on approximating a linearization of the underlying system. 
NSDMD leverages knowledge of the structure of the Navier-Stokes equations to formulate a modified DMD regression problem designed to ensure that the result is an approximation of the mean-flow-linearized operator, which is critical to extracting physical insights from any ensuing analysis. Specifically, the method uses data snapshots of velocity fluctuations and of the nonlinear forcing (in the time domain) to explicitly isolate the action of the mean-flow-linearized operator on the velocity field. NSDMD then finds a best-fit linear operator that maps the state (velocity perturbations) to its time derivatives with the nonlinear terms subtracted out, as shown schematically in Figure~\ref{fig_nsdmd}.

The remainder of this article is organized as follows. We begin with the mathematical formulation of the proposed method in section~\ref{sec:method}, describe the numerical examples used to demonstrate it in~\ref{sec:examples}, present and discuss our results in~\ref{sec:results}, and offer our conclusions in~\ref{sec:conclusions}.





\section{Proposed method\label{sec:method}}

Here we introduce a new data-driven modal decomposition that is the main contribution of this work. We present a generic formulation of the method for nonlinear dynamical systems, followed by a discussion on its application to turbulent flows. We close this section with practical implementation details for large datasets.

 \subsection{Nonlinearity-subtracted dynamic mode decomposition}

As a starting point, we take a generic nonlinear dynamical system 
\begin{equation}
    \b{\dot{x}}=\b{Ax} + \b{f}(\b{x}),\label{dynamics}
\end{equation}
where the overdot denotes time-differentiation, $\b{x}\in\mathbb{C}^n$ is the state of the system, $\b{A}$ is the operator governing the linear part of the dynamics and $\b{f}: \mathbb{C}^n \rightarrow \mathbb{C}^n$ corresponds to the purely nonlinear contribution to the dynamics. We assume that a numerical representation is available for the nonlinear part of the dynamics but unavailable for the linear part. Our aim is then to identify a low-rank approximation of $\b{A}$ from time series measurements of the state.

Given a set of $m$ measurements of the state $\b{x}_j=\b{x}(t_j)$ for $j= 1,\dots,m$, which may be acquired from one or several trajectories, we may assemble the data matrices
\begin{equation*}
    \b{X} = \left[\b{x}_1 \  \ \b{x}_2 \ \cdots \ \b{x}_m\right] \in\mathbb{C}^{m\times n}\quad \text{and } \quad
    \b{Y} = \left[\b{\dot{x}}_1\!-\!\b{f}_1 \ \ \b{\dot{x}}_2\!-\!\b{f}_2 \ \cdots \ \b{\dot{x}}_m\!-\!\b{f}_m\right]  \in\mathbb{C}^{m\times n},
\end{equation*}
where  $\b{f}_j=\b{f}(\b{x}_j)$ is the nonlinearity acting on $\b{x}_j$, and $\b{\dot{x}}_j=\b{\dot{x}}(t_j)$ are the time derivatives that may be approximated from the sequential state data, via finite differences, for example, if not directly measured. We may now formulate an optimization problem to identify a rank-$r$ approximation of $\b{A}$
\begin{equation}
  \b{A} = \underset{\mathrm{rank}(\b{A})\le r}{\mathrm{argmin}} \ \| \b{Y}  -\b{A} {\b{X}} \|_F,
  \label{opt_nsdmd}
\end{equation}
which is very similar to the DMD problem~\citep{schmid2010jfm}. Importantly, the original DMD algorithm only approximates the solution to Eq.~\eqref{opt_nsdmd}, and the exact closed-form solution is obtained using the low-rank DMD algorithm introduced by~\cite{heas2022jns}. In the context of NSDMD, using the optimal low-rank solution makes a significant difference because, in this scenario, the columns of $\b{X}$ and the columns of $\b{Y}$ can span very different subspaces, as opposed to what happens for DMD problems, where they span similar subspaces.

As in~\cite{heas2022jns}, the optimal solution to Eq.~\eqref{opt_nsdmd} is given by
\begin{equation}
    \b{A}=\b{U_Z U}\T_{\b{Z}} \b{Y}\b{X}^\dagger,\label{lrdmd}
\end{equation}
where $H$ denotes the Hermitian transpose, $\dagger$ the Moore-Penrose pseudoinverse and $\b{U_Z}\in\mathbb{C}^{n\times r}$ corresponds to the leading-$r$ left singular vectors of $\b{Z}=\b{Y}\b{V_X V}\T_{\b{X}}$, with $\b{V_X}\in\mathbb{C}^{m\times m}$ the right singular vectors of $\b{X}$. For the latter, we assumed $m<n$, which is the typical scenario encountered in fluid dynamics. We point out that the expression in Eq.~\eqref{lrdmd} results in a large $n\times n$ matrix that is never used in practice. Instead, a much smaller $r \times r$ matrix is built by projecting onto $\b{U_Z}$
\begin{equation}
    \b{\hat{A}}=\b{U}\T_{\b{Z}} \b{A}\b{U_Z}=\b{U}\T_{\b{Z}} \b{Y}\b{X}^\dagger\b{U_Z}.\label{Ahat}
\end{equation}
Importantly, it can be shown that $\b{A}$ and $\b{\hat{A}}$ share the same non-zero eigenvalues, and the associated eigenvectors $\b{v}_j\in\mathbb{C}^{n}$ and $\b{\hat{v}}_j\in\mathbb{C}^{r}$ for $j=1,\dots,r$, respectively, are related via the change of coordinates $\b{v}_j=\b{U_Z}\b{\hat{v}}_j$. We refer the readers to~\cite{heas2022jns} for a proof. Because $\b{\hat{A}}$ is an $r \times r$ matrix, its eigendecomposition can typically be performed using direct solvers. Additionally, the operator can be used to perform other linear analyses, such as DDRA~\citep{herrmann2021jfm}.

Moreover, if the system of interest is known to be shift-equivariant, as is the case for fluid flows with homogeneous directions, then, for the identified model to preserve this property, $\b{A}$ is required to be a circulant (or block-circulant) matrix~\citep{baddoo2023prsa}. The optimization problem [Eq.~\eqref{opt_nsdmd}] can then be modified to constrain the solution space to only matrices that possess that structure, and its solution can be obtained using piDMD~\citep{baddoo2023prsa}. Fortunately, by applying a Fourier transform to the data in the homogeneous directions, the optimization problem for the shift-equivariant case reduces to a series of uncoupled problems of the same form as Eq.~\eqref{opt_nsdmd}, each with the solution given by Eq.~\eqref{lrdmd}, for every individual wavenumber (or wavenumber combination for multiple homogeneous directions).

Another remark of practical relevance is that the pseudoinverse of $\b{X}$ appearing in Eq.~\eqref{Ahat} may be ill-conditioned; it is often approximated using a truncated singular value decomposition (SVD) instead of a complete one. This truncation should not be very aggressive, as its only aim is to improve the conditioning of the problem.


\subsection{NSDMD for incompressible turbulent flows}

We now consider the formulation of NSDMD for the specific case of incompressible turbulent flows. For a statistically stationary flow with mean velocity and pressure fields $\b{\overline{u}}$ and $\overline{p}$, respectively, the dynamics of the velocity and pressure fluctuation fields $\b{u'}$ and $p'$ are governed by
\begin{equation}
    \d{\b{u'}} + \b{u'}\cdot \nabla\b{u'} - \overline{(\b{u'}\cdot \nabla)\b{u'}} = \b{\mathcal{L}}\b{u'} -\nabla p' \quad \text{and } \quad \nabla \cdot \b{u'} = \b{0}\label{perturb_eqs}
\end{equation}
completed by suitable boundary and initial conditions, where $\overline{ \ \!\cdot \ \!}$ denotes an ensemble-averaging operation (or time averaging in the ergodic case), and the linear operator acting on the perturbations, $\b{\mathcal{L}}\b{u'}=Re^{-1}\nabla^2 \b{u'} - \b{u'}\cdot \nabla\b{\overline{u}} - \b{\overline{u}}\cdot \nabla\b{u'}$, corresponds to the mean-flow-linearized operator.
If Eq.~\eqref{perturb_eqs} is projected onto a divergence-free basis, then the fluctuating pressure gradient may be eliminated, provided that the system has boundary conditions that are periodic or have a zero component normal to the boundary~\citep{rowley2017arfm}.

To apply the proposed NSDMD method, we can recognize discretized versions of the terms in Eq.~\eqref{perturb_eqs} with those in Eq.~\eqref{dynamics}. Specifically, if we take $\b{u'}$ as our state $\b{x}$, then the term $\b{u'}\cdot \nabla \b{u'}$, referred to as nonlinear forcing in the resolvent framework (but here in the time domain), becomes our nonlinearity $\b{f}(\b{x})$. Moreover, if we project onto divergence-free modes, we eliminate the pressure and continuity is automatically satisfied. Therefore, an equation of the form of Eq.~\eqref{dynamics} governs the evolution of the discretized divergence-free velocity fluctuations, where the matrix $\b{A}$ that we want to approximate from data using NSDMD corresponds to a discretization of $\b{\mathcal{L}}$.

Application of the method then requires collecting a time sequence of snapshots of the velocity fluctuation field to build the data matrix $\b{X}$. Additionally, the $\b{Y}$ data matrix requires snapshots of $\partial\b{u'}/\partial t$ that may be approximated from samples of $\b{u'}$ using finite differences, and snapshots of the nonlinear forcing that may be outputted by the same code being used to generate the simulation data, or also approximated numerically from samples of $\b{u'}$ if the data is on a structured grid. Keep in mind that a well-converged mean flow is also needed to appropriately define the velocity perturbations. Finally, cell volumes are also required to be used as inner-product weights.


\subsection{Dealing with large datasets}

When dealing with large datasets of high-dimensional systems such as turbulent flows, the data matrices $\b{X}$ and $\b{Y}$, of size $n \times m$ with $n \gg m$, usually become too big to reside in memory. Fortunately, this limitation can be circumvented by loading the data sequentially and using the method of snapshots~\citep{sirovich1987qam} to compute the correlation matrices $\b{C_{XX}} = \b{X}\T\b{X}$, $\b{C_{XY}} = \b{X}\T\b{Y}$ and $\b{C_{YY}} = \b{Y}\T\b{Y}$. Once these $m\times m$ matrices have been computed, they can be used to build $\b{\hat{A}}$ as
\begin{equation}
    \b{\hat{A}}=\b{\Sigma_Z V}\T_{\b{Z}} \b{\Sigma}^{-2}_{\b{X}}\b{V}\T_{\b{X}}\b{C_{XY}V_X V_Z \Sigma}^{-1}_{\b{Z}},\label{Ahat_corr}
\end{equation}
where $\b{V_X}$ corresponds to the matrix of right singular vectors of $\b{X}$, which are more conveniently computed as the eigenvectors of $\b{C_{XX}}$, and $\b{V_Z}$ and $\b{\Sigma_Z}$ correspond to the right singular vectors and the singular values of $\b{Z}$, which are more conveniently computed as the eigenvectors and the square root of the eigenvalues of $\b{V}\T_{\b{X}}\b{C_{YY}}\b{V_X}$, respectively. Note that the $\b{\hat{A}}$ resulting from Eqs.~\eqref{Ahat} and~\eqref{Ahat_corr} are equivalent. However, using the latter avoids operating on any vectors of size $n$ after the correlation matrices have been computed; therefore, it can be significantly reduce the computational effort and memory allocation cost of the method.

\section{Numerical examples and datasets\label{sec:examples}}

\begin{figure}
    \centering
    \includegraphics[width=1\linewidth]{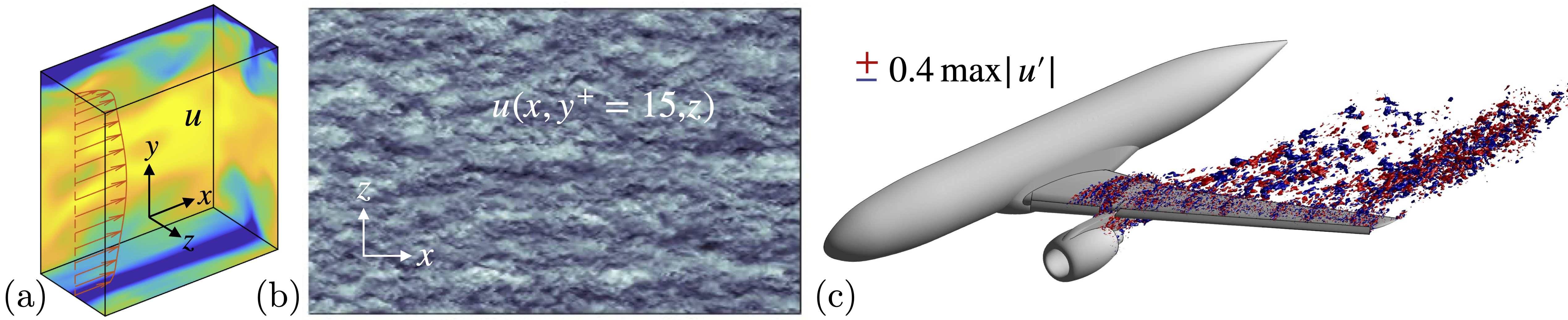}
    \caption{Turbulent flow examples used to demonstrate NSDMD. (a) Minimal channel flow data from~\cite{herrmann2023jfm}. (b) ZPGFP boundary-layer data from~\cite{towne2023aiaa}. (c) Flow over JSM aircraft, numerical setup from~\cite{goc2021flow}.}
    \label{datasets}
\end{figure}

Here we describe the numerical setups and generated datasets for the three examples studied that are shown schematically in Figure~\ref{datasets}.

\subsection{Minimal channel flow}

For a first example, we use a minimal channel flow~\citep{jimenez1991jfm} at $Re_{\tau}=185$ with the same setup as in~\cite{herrmann2023jfm}. We use the spectral code Channelflow \citep{gibson2014chflow} to perform DNSs and generate a dataset comprised of a long sequence of snapshots acquired after statistical stationarity is reached. Specifically, $75,000$ velocity field snapshots $\b{x}_j$ are saved every $1.63$ viscous inner time units and spanning $661$ eddy-turnover times. The streamwise velocity for a typical flow field snapshot is shown in Figure~\ref{datasets}(a). Snapshots for the time derivative of the state $\b{\dot{x}}_j$ are approximated using an eigth-order central finite difference scheme, and snapshots for the nonlinearity $\b{f}_j$ are computed using Chebyshev and Fourier spectral differentiation. After the time derivatives are computed, we down-sample the data in time, retaining $1$ of every $20$ snapshots, yielding a final total of $m=3750$ snapshots.

After Fourier transforming in the horizontal directions, an independent analysis can be carried out for each pair of streamwise and spanwise wavenumbers. This ensures that the resulting model respects the shift-equivariance of the flow in the homogeneous directions~\citep{baddoo2023prsa}. Moreover, discrete symmetries inherent to the geometry of the problem are also incorporated by augmenting the dataset with the action of the appropriate transformations on the original snapshots, multiplying the total number of snapshots by four.

Furthermore, to compare against the results based on NSDMD, we perform resolvent analysis based on the governing equations linearized about the mean flow. The mean flow is computed from the DNS snapshots and used to build the mean-flow-linearized operator with an in-house code based on the Orr-Sommerfeld/Squire formulation. Our code uses Chebyshev spectral collocation to discretize the wall-normal direction with the same grid used in the DNS.

\subsection{Zero-pressure-gradient flat-plate boundary-layer flow}

The time-resolved velocity fields of the BL1 zero-pressure-gradient flat-plate (ZPGFP) turbulent boundary-layer dataset from \cite{towne2023aiaa} are used. These were obtained from a DNS of the incompressible Navier-Stokes equations using a staggered second-order central finite difference scheme for spatial discretization, and a third-order Runge-Kutta scheme combined with the fractional-step method for time advancement.

We use a subset of the full computational domain in the streamwise direction, covering from $Re_{\tau} \approx 559–636$. The dataset contains $10,000$ snapshots acquired every $\Delta t^+ \approx 1.5$ and spanning $26$ eddy-turnover times. A snapshot of the streamwise velocity field in a wall-normal plane at $y^+=15$ is shown in Figure~\ref{datasets}(b). The mean flow field is subtracted to obtain the velocity perturbation snapshots $\b{x}_j$; the nonlinear forcing snapshots $\b{f}_j$ are computed from the same data using second-order central finite differences to build the velocity gradients, and time derivatives $\b{\dot{x}}_j$ are approximated using an eigth-order central finite difference scheme. Moreover, reflection symmetry along the midplane perpendicular to the spanwise direction ($z$) is exploited to augment the dataset, doubling the amount of snapshots. Snapshots are then Fourier transformed in $z$, and the analysis is carried out independently for individual wavenumbers.

\subsection{Flow over the JAXA standard model aircraft}

Wall-modeled LESs (WMLESs) of the compressible flow over the Japanese Aerospace Exploration Agency (JAXA) standard model aircraft (JSM) are performed using charLES~\citep{bres2018large,goc2021flow}. In particular, we consider the same numerical setup as in~\cite{goc2021flow} to simulate the half-body JSM with nacelle inside a low-speed wind tunnel. We use a mesh with 41 million cells and select an angle of attack $\alpha = 4.0^\circ$, a free stream Mach number $M_\infty = 0.172$ and a Reynolds number $Re_c = 1.93\cdot10^6$, where we use the mean aerodynamic chord $L_c$ and free stream velocity $U_\infty$ as the characteristic length and velocity scales, respectively.

The simulation is initially run for $60$ time units, based on $L_c/U_\infty$ as the timescale, which is sufficient for transients due to the initial condition to die out. A subsequent simulation is performed over an additional ten time units to compute the mean velocity field used to define perturbations. Finally, a third simulation is run to collect data for the DMD and NSDMD analyses. Because the flow is at a low Mach number, we define our state using only the velocity perturbation field and disregard the pressure, as if it were an incompressible flow. Snapshots of the perturbations $\b{\dot{x}}_j$ and the nonlinear forcing $\b{f}_j$, which is computed using the exact flux operators implemented in the code, are registered every $0.112$ time units over a window of $56$ time units, for a total of $m=500$ snapshots. Additionally, the time derivatives of the state $\b{\dot{x}}_j$ at the acquisition times $t_j$ are approximated using a fourth-order central finite difference scheme, which uses the velocity fields at two preceding and two succeeding time steps, with $\Delta t=1\cdot10^{-6}$. Importantly, only a small region of the computational domain that surrounds the aircraft wing and its near wake is used for the analysis. A typical snapshot of streamwise velocity perturbations on this smaller region is shown in Figure~\ref{datasets}(c).

\section{Results and discussion\label{sec:results}}

\subsection{Data requirements}

\begin{figure}
    \centering
    \includegraphics[width=1\linewidth]{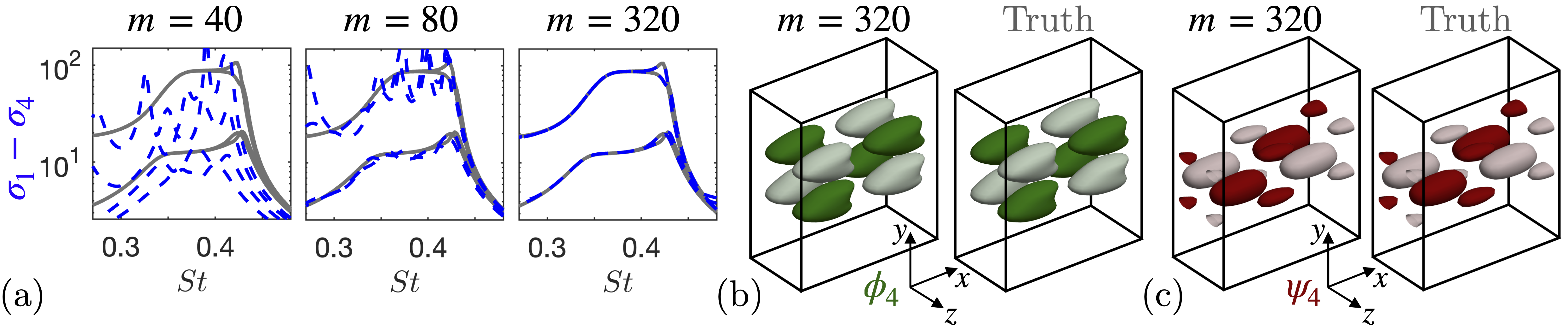}
    \caption{NSDMD-based DDRA of minimal channel flow for $k_x=2\pi/L_x$ and $k_z=2\pi/L_z$. (a) Convergence of the leading four resolvent gains with the number of snapshots $m$. Data-driven and equation-based gains are shown in dashed blue and solid gray lines, respectively. (b) Isosurfaces of the wall-normal component of $\b{\phi}_4$. (c) Isosurfaces of the streamwise component of $\b{\psi}_4$. Modes are shown for their most amplified frequency, isosurfaces correspond to $\pm 0.4$ of their peak magnitude and real left- and right-going waves are added together.}
    \label{fig_chflow}
\end{figure}

Here we assess the effect of the number of snapshots on the results produced by NSDMD on the minimal channel flow DNS data. Specifically, we look at the leading and suboptimal resolvent gains and modes for streamwise and spanwise wavenumbers $k_x=2\pi/L_x$ and $k_z=2\pi/L_z$, respectively. We then compare the results of equation-based resolvent analysis, based on the mean flow from the DNS, with those from DDRA based on an NSDMD model trained on $m=40$, $80$ and $320$ snapshots, as shown in Figure~\ref{fig_chflow}. In fact, these are just $10$, $20$ and $40$ snapshots from the DNS that are augmented considering the discrete symmetries of the problem, leading to four times the respective amounts. We find that a total of $m=320$ snapshots is enough for the DDRA to converge to the equation-based results, not only for the leading gain distribution and forcing and response modes, but also up to the fourth suboptimal gains and modes, as shown in Figure~\ref{fig_chflow}. The small amount of data required in this case is explained by the small underlying dimension of the system, which, considering our discretization of $101$ points in the wall-normal direction, is $n=297$ (excluding boundary points). Note that, dealing with a separate problem for each streamwise and spanwise wavenumber combination is critical for the success of the method with a small number of snapshots in this case. This accounts for the shift-equivariance of the system in the homogeneous directions, significantly reducing the number of free parameters that are being fit because the dynamics are forced to be decoupled between wavenumber tuples, which drastically reduces the data requirements~\citep{baddoo2023prsa}. 

\subsection{Implementation for large datasets}

\begin{figure}
    \centering
    \includegraphics[width=1\linewidth]{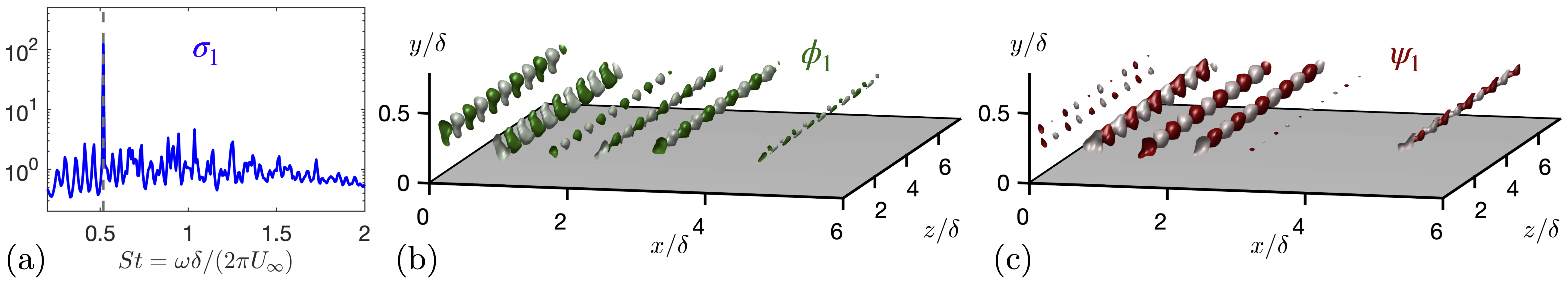}
    \caption{NSDMD-based DDRA of the ZPGFP boundary-layer flow for $k_z=10\pi/L_z$. (a) Leading resolvent gain as a function of $St$. Leading (b) forcing and (c) response modes at the most amplified frequency. Modes are shown as positive (light) and negative (dark) isosurfaces of the streamwise components at $0.75$ of their peak magnitudes. Real left- and right-going waves are added together, and only the upstream half of the domain, where mode activity is concentrated, is shown.}
    \label{fig_bl}
\end{figure}

We test our scalable NSDMD implementation, that relies on Eq.~\eqref{Ahat_corr} rather than Eq.~\eqref{Ahat}, to produce a low-rank approximation of the mean-flow-linearized operator of a turbulent boundary-layer from DNS data. Specifically, we use the Fourier-transformed snapshots corresponding to the most energetic spanwise wavenumber, $k_z=10\pi/L_z$, from the ZPGFP boundary-layer dataset described in the previous section. Even though we work with a single wavenumber, the state dimension for the problem is still $n\sim 10^5$, which, combined with the large number of snapshots $m\sim 10^4$, starts demanding considerable resources in terms of memory allocation and compute time. 

Before applying NSDMD, we truncate the SVD of $\b{X}$ based on the optimal hard threshold for singular values by~\cite{gavish2014ieeetit}. Subsequently, we use NSDMD to obtain a rank $r=400$ model, truncating the SVD of $\b{Z}$ in Eq.~\eqref{Ahat_corr}, and use it to perform DDRA. We compute the leading resolvent gains and the forcing and response modes at the most amplified frequency, as shown in Figure~\ref{fig_bl}. Our scalable NSDMD implementation successfully produces familiar looking modes, with backwards-tilting forcing structures concentrated upstream of the forward-tilting response structures, while circumventing memory allocation limitations and avoiding unnecessary and redundant computations.

\subsection{Comparison of NSDMD with DMD}

We apply DMD and NSDMD to the large dataset for the flow over the JSM. Because of the relatively low number of snapshots, $m=500$, we do not rank-truncate the SVDs of the data matrices. Note that here DMD is performed in continuous time, meaning that we use the time derivatives of the state rather than the time-advanced state in the data matrix $\b{Y}$ for DMD. Other than that, we use the original formulation from~\cite{schmid2010jfm}, resulting in modes that live in the column space of $\b{X}$. Therefore, the main methodological differences between the two approaches are that, for NSDMD, the nonlinear forcing is subtracted out from the time derivatives to form the data matrix $\b{Y}$, and the optimal low-rank solution to the optimization problem Eq.~\eqref{opt_nsdmd} is used.

\begin{figure}
    \centering
    \includegraphics[width=1\linewidth]{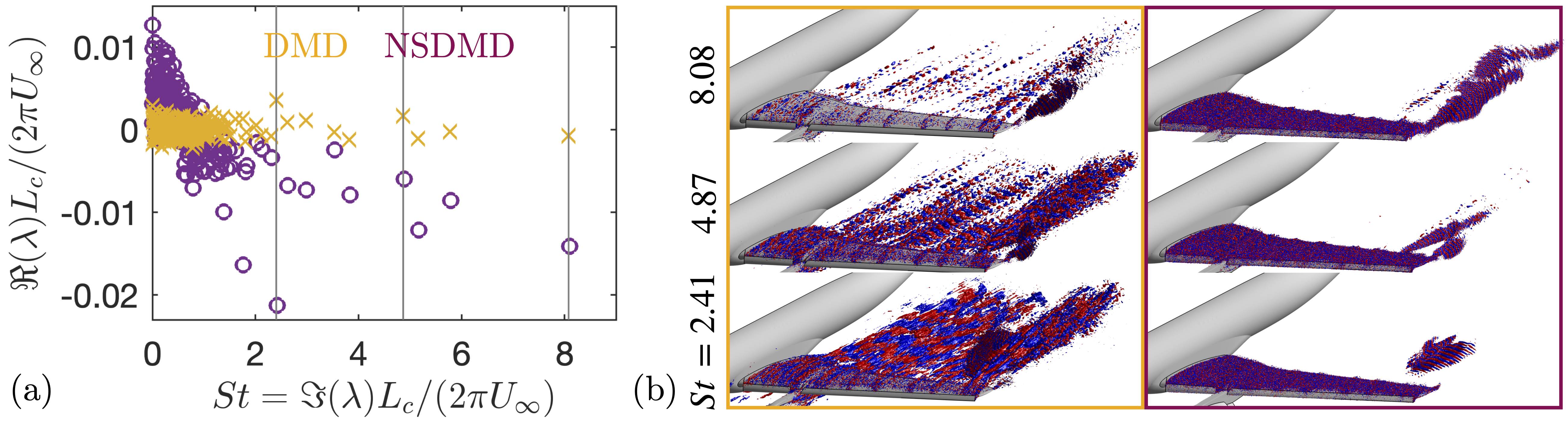}
    \caption{Comparison between DMD and NSDMD (a) eigenvalues and (b) modes. DMD modes (left panel) and NSDMD modes (right panel) are shown as isosurfaces of the real parts of their streamwise components at $\pm 0.05$ and $0.005$ of their peak magnitudes, respectively.}
    \label{fig_jsm}
\end{figure}

We find that DMD and NSDMD eigenvalues share the same imaginary parts, while their real parts differ, as shown in Figure~\ref{fig_jsm}(a). Real parts of DMD eigenvalues are all close to zero, as is expected from the analysis of any statistically stationary turbulent flows. However, by accounting for the nonlinear forcing, NSDMD eigenvalues with high frequencies are pushed down in the complex plane, while the ones with lower frequencies are pulled up. Furthermore, in our experimentation with NSDMD, we have consistently found that higher frequency modes are less sensitive to methodological details, such as the amount of data snapshots and rank truncation, whereas we tend to accumulate spurious modes at the lower frequencies. Improving the robustness of low-frequency NSDMD modes is an important challenge that should be addressed in future research.

Moreover, we find significant differences in the mode shapes obtained with both methods, as shown in Figure~\ref{fig_jsm}(b). This can be better understood in terms of the subspaces generated by the modes. With the implementations used here, DMD modes live in the span of the dominant POD modes of $\b{X}$, whereas NSDMD modes live in the span of the dominant POD modes of $\b{Z}=\b{YV_{X}V}\T_{\b{X}}$, that is, the POD modes of $\b{Y}$ after projecting its rows onto $\b{V_X}$. The latter can be interpreted as the dominant structures in $\b{\dot{x}}_j-\b{f}_j$ whose time evolution is correlated with the dynamics of $\b{x}_j$. Importantly, the spatial footprint of NSDMD modes reveal the region of influence of the mean-flow-linearized operator when acting on realizable velocity fluctuations in the statistically stationary turbulent flow. For the flow over the JSM, this region of influence concentrates on the wing surface boundary-layers and the wing-tip vortex, while excluding wake structures.

\section{Conclusions\label{sec:conclusions}}
We developed nonlinearity-subtracted dynamic mode decomposition (NSDMD) — a novel data-driven method that aims at enabling physics learning from large numerical datasets of industrially relevant turbulent flows. The method leverages knowledge of the structure of the Navier-Stokes equations to approximate the mean-flow-linearized operator from time-resolved flow field and nonlinear forcing data snapshots. The learned operator allows linear analyses of turbulent flows to be performed as a post-processing step on simulation data obtained with any available high-fidelity CFD code. For statistically stationary flows, the sensitivity of the ensuing results to the amount of data snapshots can be used to assess their convergence. Moreover, the resulting NSDMD modes identify the regions in space where linear mechanisms are active through the action of the mean-flow-linearized operator on realizable velocity fluctuations for the underlying turbulent flow. We suggest that future work should focus on filtering spurious eigenvalues that are typically found for low-frequency modes.\\

\section*{Acknowledgments}
We are very grateful to Prof. Parviz Moin and Dr. Salvador Gomez for their support and hospitality during the CTR Summer Program 2024, and to Yongyun Hwang and Ahmed Elnahhas for helpful comments and discussions. 



\bibliographystyle{ctr}


\end{document}